# N-Graphdiyne as a Tunable Platform for Stabilizing Light Metals toward High-Capacity Reversible Hydrogen Storage

Wael Othman[1,2], Ibrahim Alghoul[3,4], Nacir Tit[3,4*], K-F. Aguey-Zinsou[5], and Tanveer Hussain[6*]

[1] Biomedical Engineering and Biotechnology, Khalifa University, Abu Dhabi, United Arab Emirates

[2] Healthcare Engineering Innovation Group (HEIG), Khalifa University, Abu Dhabi, United Arab Emirates

[3] Physics Department, United Arab Emirates University, Al Ain, United Arab Emirates

[4] Water Research Center, United Arab Emirates University, Al Ain, United Arab Emirates

[5] MERLin, School of Chemistry, University of Sydney, NSW 2006, Australia

[6] School of Science and Technology, University of New England, Armidale, New South Wales, 2351, Australia

* Correspondence: ntit@uaeu.ac.ae & tanveer.hussain@une.edu.au

## Abstract:

Hydrogen ($H_2$) is a promising carbon-neutral energy carrier; however, its deployment is limited by the lack of lightweight, reversible storage media that operate under practical conditions. Here, we establish nitrogen-doped graphdiyne (N-GDY) as a programmable two-dimensional platform for stabilizing dispersed light-metal dopants and enabling high-capacity molecular $H_2$ physisorption. The computational package involves density functional theory (DFT) combined with ab initio molecular dynamics (AIMD) and Langmuir-based statistical thermodynamic modeling. The results revealed that N-sites of N-GDY bind up to five Li, Na, K, and Ca atoms per primitive cell with binding energies of −2.27, −1.57, -1.80, and −2.13 eV, respectively, exceeding their respective bulk cohesive energies. AIMD simulations at 400 K further confirm the structural robustness of the decorated frameworks and the absence of metal aggregation. The polarised metal centres activate reversible $H_2$ adsorption through electrostatic and dispersion interactions, with average adsorption energies falling within the optimal window (−0.15 to −0.35 eV per $H_2$). Sequential adsorption analysis reveals uptake of up to 25 $H_2$ molecules per primitive cell, achieving intrinsic gravimetric capacities of 13.08, 10.82, 9.23, and 9.15 wt% for Li-, Na-, K-, and Ca-functionalized systems, respectively. Thermodynamic analysis indicates favorable adsorption–desorption behavior under near-ambient conditions, with Li- and Ca-functionalized systems exceeding the 6.5 wt% U.S. Department of Energy's ultimate system-level target when considering intrinsic material capacity. These results identify N-GDY as a chemically tunable scaffold for dispersing lightweight metals and provide a mechanistic design strategy for achieving high-capacity, reversible hydrogen storage in porous two-dimensional materials.

**Keywords:** Hydrogen storage; N-Graphdiyne; Metal decoration; DFT; AIMD; Thermodynamics

## I. Introduction

The global shift away from fossil fuels is no longer just an environmental preference but a strategic necessity to combat the dual threats of climate change and resource depletion. A diversified portfolio of low-carbon energy sources, particularly the virtually inexhaustible sources of solar and wind power, has proven both economically viable and environmentally sustainable, enabling the decoupling of economic growth from carbon emissions [1]. However, the widespread adoption of these renewable technologies requires reliable and efficient energy storage systems, which have traditionally been dominated by batteries that are currently facing material resource availability challenges. Recently, hydrogen ($H_2$) has emerged as a promising energy carrier and alternative energy storage approach due to its sustainability and eco-friendliness [2].

Despite these advantages, the practical implementation of $H_2$ energy is limited by challenges associated with storage efficiency. Although $H_2$ possesses an exceptionally high gravimetric energy density, its volumetric density under ambient conditions is extremely low. For instance, liquid $H_2$ has a density of only 71 kg/m$^3$, which is approximately 14 times smaller than that of water, and requires cryogenic temperatures ($T_b$ = -253 °C = 20 K) for liquification [3]. Consequently, conventional $H_2$ storage methods are based on high-pressure gas cylinders or cryogenic liquid hydrogen that involve significant engineering complexity, energy losses, and safety concerns.

To address these limitations, solid-state $H_2$ storage materials have emerged as an alternative approach capable of enabling $H_2$ storage under moderate temperatures and pressures [4]. In these systems, $H_2$ is either (i) physically adsorbed (physisorption) onto high-surface-area materials such as metal-organic frameworks (MOFs) and porous carbons, or (ii) chemically bonded (chemisorption) within the crystal lattice of metal hydrides and complex hydrides [5]. These approaches enable $H_2$ storage at relatively high densities under safer, near-ambient conditions, with physisorption-based systems being more favorable because it offers reversible $H_2$ uptake while avoiding the high enthalpy requirements associated with chemical hydride formation. When coupled with a Proton Exchange Membrane (PEM) fuel cell, the stored $H_2$ can be released on demand and react with atmospheric oxygen to generate electricity with zero localized carbon emissions, making it particularly attractive for automotive and portable energy applications [6].

In recent years, two-dimensional (2D) materials have attracted significant attention owing to their large surface areas, tunable electronic properties, and structural flexibility. The discovery of graphene by Geim and Novoselov in 2004 has not only earned them the 2010 Nobel Prize in Physics but also stimulated extensive exploration of 2D materials [7-9]. Since then, numerous atomically thin systems, including nitride [10] and oxide monolayers [11-16], have been identified and investigated for potential applications in energy storage, catalysis, electronics, and gas sensing. Among these materials, nitrogen-containing carbon frameworks have gained considerable interest due to their enhanced chemical reactivity and their ability to stabilize metal atoms through strong coordination interactions [17]. Porous carbon nitrides and $C_xN_{1-x}$ monolayers have been predicted and synthesized, exhibiting tunable structural and electronic properties that are particularly suitable for energy-related applications [18]. Cai et al. [19] performed a systematic density functional theory (DFT) investigation of $C_xN_{1-x}$ monolayers, including $C_3N_5$, $C_3N_4$, CN, $C_2N$, and $C_3N$, revealing that increasing nitrogen (i.e., $1-x > 0.33 = 33\%$) content promotes the formation of porous structures with favorable thermodynamic stability.

Another emerging member of the porous carbon family is graphdiyne (GDY), a two-dimensional carbon allotrope composed of sp- and $sp^2$-hybridized carbon atoms connected through diacetylene bridges [20]. The intrinsic porosity and extended π-conjugation of GDY distinguish it from graphene and make it particularly attractive for adsorption-based applications. Recently, nitrogen-doped graphdiyne (N-GDY) monolayers have been synthesized through interfacial coupling reactions, enabling the formation of well-defined conjugated networks such as $C_{18}N_6$, $C_{24}N_4$, and $C_{36}N_6$ [21, 22]. These materials exhibit high structural stability and tunable electronic properties, and have already demonstrated promising performance in electrochemical applications, including oxygen reduction reactions and alkali-ion storage [23, 24].

Despite these advances, the potential of N-GDY as a host material for $H_2$ storage remains largely unexplored. In particular, the presence of nitrogen atoms introduces localized coordination sites capable of stabilizing dispersed metal atoms, which can polarize $H_2$ molecules and induce favorable electrostatic and van der Waals interactions that enable reversible physisorption. For practical $H_2$ storage applications, the adsorption energy of $H_2$ should ideally fall within the range of -0.20 to -0.60 eV per molecule to ensure reversible adsorption–desorption behavior under near-ambient conditions. In this context, the U.S. Department of Energy (DOE) has established an ultimate system-level gravimetric capacity target of 6.5 wt% for onboard $H_2$ storage systems [25].

In the present work, we investigate the $H_2$ storage capability of N-GDY (i.e., $C_{18}N_6$) monolayer functionalized with light metals (Li, Na, K, and Ca). Using DFT combined with ab initio molecular dynamics (AIMD) simulations and statistical thermodynamic analysis, we systematically evaluate the stability of light-metal functionalization, the absence of metal aggregation relative to bulk cohesive energies, and the $H_2$ adsorption behavior of the functionalized systems. Furthermore, we determine the maximum $H_2$ uptake capacity and evaluate the corresponding gravimetric storage performance under practical adsorption–desorption conditions. The present study provides fundamental insights into the role of light-metal decoration in enhancing $H_2$ adsorption and highlights N-GDY as a promising candidate material for high-capacity reversible $H_2$ storage.

## II. Computational Methods

DFT calculations were performed using the Vienna Ab-initio Simulation Package (VASP) [26]. The interaction between valence electrons and ionic cores was described using the projector augmented-wave (PAW) method, while the electronic wave functions were expanded in a plane-wave basis set with an energy cutoff of 500 eV [27]. The exchange-correlation interaction was treated using the general gradient approximation (GGA) within the Perdew-Burke-Ernzerhof (PBE) hybrid functional form [28, 29]. To account for long-range dispersion interactions (van der Waals), the DFT-D3 correction with Grimme's scheme was employed [30]. The Brillouin zone was sampled using a Monkhorst–Pack k-point mesh of 23 × 23 × 1 [31]. Structural optimizations were performed until the total energy converged within $10^{-6}$ eV, and the atomic forces were reduced below 0.01 eV/Å. Charge transfer between the light-metal adatoms and the substrate was analyzed using Bader charge analysis, and the corresponding charge density difference (CDD) was calculated to visualize the redistribution of electronic charge after functionalization [32].

The average binding energy ($E_{bind}$) of a light-metal ($M$) atom embedded in the pores of N-GDY monolayer is calculated as:

$$E_{bind} = \frac{E_{N-GDY+nM} - (E_{N-GDY} + n \cdot E_M)}{n} \quad (1),$$

where $E_{N-GDY+nM}$, $E_{N-GDY}$, and $E_M$ represent the total energies of the N-GDY decorated with *n* metal atoms, the pristine N-GDY monolayer, and an isolated metal atom, respectively. The calculated average $E_{bind}$ are compared with the cohesive energies of the corresponding bulk metals to evaluate the stability of the embedded metal atoms and their resistance to clustering. To further assess the thermal stability, ab-initio molecular dynamics (AIMD) simulations are performed at 400 K using the Nosé–Hoover thermostat. The simulations were conducted with a time step of 1 fs for a total duration of 5 ps.

The average adsorption energy ($E_{ads}$) of $H_2$ molecules on a metal-decorated N-GDY is calculated using:

$$E_{ads} = \frac{E_{N-GDY+mH_2} - (E_{N-GDY} + m \cdot E_{H_2})}{m} \quad (2),$$

where, $E_{N-GDY+mH_2}$, $E_{N-GDY}$, and $E_{H_2}$ denote the total energies of the hydrogenated system with *m* $H_2$ molecules, the bare system, and the isolated $H_2$ molecule, respectively.

The theoretical gravimetric $H_2$ storage capacity ($C_T$) is defined as:

$$C_T(wt\%) = \left[\frac{N_T \cdot M(\mathrm{H_2})}{N_T \cdot M(\mathrm{H_2}) + M(Sub)}\right] \times 100\% \quad (3),$$

where $M(\mathrm{H_2})$ and $M(\mathrm{Sub})$ represent the masses of the $H_2$ molecule and the substrate, respectively, and $N_T$ denotes the maximum number of adsorbed $H_2$ molecules.

To evaluate the adsorption–desorption behavior under practical operating conditions, a statistical thermodynamic analysis [33, 34] is carried out based on the grand canonical partition function ($z$):

$$z = 1 + \sum_{i=1}^{n} e^{-\frac{(E_b^i - \mu)}{k_B T}} \quad (4),$$

where $n$ is the maximum number of adsorbed $H_2$ molecules, $E_b^i$ is the adsorption energy of the $i$-th adsorbed $H_2$ molecule, $k_B$ is the Boltzmann constant (1.38 x $10^{-23}$ J/K), and $T$ is the absolute temperature. The chemical potential of $H_2$ gas, $\mu$, is expressed as a function of $P$ and $T$ as:

$$\mu_{H_2}(P,T) = \Delta H + T\Delta S + k_B T ln \frac{P}{P_0} \quad (5),$$

where $\Delta H$ and $\Delta S$ represent the enthalpy and entropy changes of $H_2$ molecules, respectively, $P$ is the $H_2$ pressure, and $P_0$ is the standard atmospheric pressure (1.01 x $10^5$ Pa). The values of $\Delta H$ + $T\Delta S$ are obtained from experimental thermidybnamic databases [35].

The number of $H_2$ molecules ($N$) stored in the host material is then given by:

$$N = N_0 \left[\frac{Z-1}{Z}\right] \quad (6),$$

where $N_0$ represents the maximum number of $H_2$ molecules adsorbed on the host medium (namely, N =$N_T$ at 0 K).

# III. Results and Discussion

## a. Pristine N-GDY

All structures investigated in this study are based on an N-GDY monolayer with chemical composition $C_{18}N_6$. The system crystallizes in a 2D triangular Bravais lattice with lattice parameters $a$ = $b$ = 16.038 Å and contains 24 atoms per primitive cell, as illustrated in Figure 1a. A vacuum spacing of 20 Å was introduced along the out-of-plane direction to eliminate spurious interactions between periodic images. The obtained lattice constant agrees well with previously reported theoretical values [24], confirming the reliability of the adopted computational setup.

The electronic properties of the pristine system were further analyzed through the spin-polarized band structure and projected density of states (PDOS), shown in Figure 1b. The Fermi level was set as the reference energy ($E_F$=0). In the band structure, the spin-up and the spin-down states are shown in solid-orange and dotted-blue curves, whereas in the PDOS, contributions from C and N atoms are shown in brown and grey curves, respectively. Within the GGA-PBE framework, the N-GDY monolayer exhibits a non-magnetic semiconducting behavior with a calculated band gap of 2.30 eV. While a more accurate bandgap can be obtained by exploring more sophisticated hybrid functionals such as HSE06, the present work primarily focuses on $H_2$ adsorption on metal-decorated N-GDY systems, which exhibit metallic behavior. Therefore, the use of the PBE functional provides a reasonable balance between computational accuracy and efficiency. The calculated band gap ($E_g$ = 2.30 eV) agrees well with the value reported by Mortazavi et al. using the same level of simulations [36].

The PDOS profiles reveal identical contributions from spin-up and spin-down states, confirming the paramagnetic nature of the pristine N-GDY monolayer. In addition, relatively flat bands are observed near the valence-band maximum (VBM) and conduction-band minimum (CBM), which can be attributed to the localized electronic states arising from the sp-hybridized carbon chains within the N-GDY framework.

After establishing the structural and electronic properties, we evaluated the interaction of a single $H_2$ molecule with the pristine N-GDY monolayer. The calculated $E_{ads}$ at three representative adsorption sites on the N-GDY surface were found to be weaker than the minimum threshold required for effective reversible storage (−0.15 eV per $H_2$), as shown in Supplementary Figure S1 (Supporting Information). The weak physisorption of $H_2$ on this substrate suggests that pristine N-GDY alone is not suitable for efficient hydrogen storage.

**b. Light Metal Functionalization**

Previous theoretical studies have demonstrated that porous N-GDY monolayers can effectively accommodate multiple light-metal atoms within their intrinsic pores. In particular, Makaremi et al. [24] reported strong binding of Li and Na atoms with $E_{bind}$ exceeding the corresponding cohesive energies, indicating that isolated metal atoms are thermodynamically stable against clustering. Motivated by these findings, we investigated the decoration of N-GDY with four light-metal atoms (Li, Na, K, and Ca) to evaluate their suitability for $H_2$ storage applications. Specifically, the embedment of light metals should make the surface of N-GDY more polar and thus appropriate to yield physisorption-type interactions with $H_2$ molecules. Stable embedment of up to five atoms per primitive cell of N-GDY was achieved. The incriminating light-metal functionalization is presented in Figure S2 (Supporting information). Figure 2 presents the optimized structures containing five metal atoms per primitive cell, namely: (a) 5Li@N-GDY, (b) 5Na@N-GDY, (c) 5K@N-GDY, and (d) 5Ca@N-GDY. Both top and side views are shown, and the structural stability at 0 K and 400 K is demonstrated.

The stability of N-GDY decorated with five metal dopants was supported by Figure 3a, which shows the calculated average $E_{bind}$ for sequential metal adsorption. In all cases, the $E_{bind}$ remain significantly larger in magnitude than the corresponding bulk cohesive energies ($|E_{bind}|>|E_{coh}|$), which are 1.63 eV, 1.11 eV, 0.93 eV, and 1.84 eV for Li, Na, K, and Ca, respectively. This indicates that metal dopants prefer binding to the N-GDY lattice rather than forming clusters. AIMD simulations performed at 400 K produced a time evolution of the total energy, which is plotted in Figure 3b. The system exhibits only small energy fluctuations (<1%), further confirming the structural robustness of the 5LM@N-GDY configurations (LM: Li, Na, K, Ca).

The electronic properties of the LM@N-GDY were further analyzed through the spin-polarized band structures and PDOS shown in Figure 4. The PDOS contributions of the embedded atoms are shown in the same colors as their color species in Figure 2 (i.e., L = green, Na = yellow, K = purple, and Ca = Blue). In contrast to the semiconducting pristine N-GDY, all decorated systems display metallic behavior, characterized by band crossings at the Fermi level and a finite density of states at $E_F$. Interestingly, decoration with relatively heavier metals (Na, K, and Ca) introduces magnetic behavior, resulting in ferromagnetic ground states with total magnetic moments of 0.93, 1.22, and 1.17 μB, respectively, whereas the Li-decorated system remains non-magnetic.

Table 1 shows the results of Bader charge transfer, Fermi level with respect to vacuum level, bandgap energy, and total magnetization in the four functionalized systems. Bader analysis results reveal significant charge transfer from the metal atoms to the N-GDY lattice. This trend is also clearly visualized in the CDD plots shown in Figure 5, where charge depletion occurs around the metal atoms and accumulation is observed along the C–N bonds. This behavior is consistent with the electronegativity difference between the metal atoms and the carbon–nitrogen framework ($\chi^{N} = 3.04 > \chi^{C} = 2.55 \gg \chi^{LM}$: $\chi^{Li} = 0.98$, $\chi^{Na} = 0.93$, $\chi^{K} = 0.82$, $\chi^{Ca} = 1.0$ Pauling), leading to electron donation from the metal atoms and the formation of polarized adsorption sites.

### c. Hydrogen Adsorption

The $H_2$ adsorption capability of the 5LM@N-GDY systems was systematically investigated. Figure 6 shows the relaxed configurations of the fully hydrogenated structures obtained after sequential adsorption of $H_2$ molecules on the four systems: (a) 5Li@N-GDY with 25 $H_2$, (b) 5Na@N-GDY with 25 $H_2$, (c) 5K@N-GDY with 25 $H_2$, and (d) 5Ca@N-GDY with 25 $H_2$ molecules.

The adsorption process was simulated by sequentially adding hydrogen molecules in steps of five molecules per primitive cell. In each step, $H_2$ molecules preferentially adsorb around the metal centers due to electrostatic polarization induced by the positively charged metal atoms.

The evolution of the average $E_{ads}$ with increasing $H_2$ loading is shown in Figure 7a. Even at the maximum loading of 25 $H_2$ molecules, the average adsorption energies remain stronger than the minimum threshold of 0.15 eV per $H_2$, indicating stable adsorption. Notably, the Ca-decorated system exhibits the strongest adsorption, with an average $E_{ads}$ of approximately 0.22 eV per $H_2$, suggesting that additional $H_2$ molecules could potentially be accommodated.

Using Equation (3), the corresponding theoretical gravimetric storage capacities were calculated and are presented in Figure 7b. The obtained values are 13.08, 10.82, 9.23, and 9.15 wt% for Li, Na, K, and Ca decorations, respectively. These intrinsic material capacities exceed the ultimate system-level target of 6.50 wt% established by the DOE, indicating the strong potential of metal-decorated N-GDY for $H_2$ storage applications.

**d. Thermodynamic Analysis**

The $H_2$ storage capacities obtained from DFT calculations correspond to the idealized 0 K limit within the Born–Oppenheimer “frozen-lattice” approximation. To evaluate the storage performance under realistic operating conditions, a statistical thermodynamic analysis based on the Langmuir adsorption model was performed using Equations (4–6). Figure 8 presents the average number of adsorbed $H_2$ molecules ($N_{ave}$) per primitive cell as a function of temperature ($T$) and pressure ($P$) for the four decorated systems. As expected, hydrogen adsorption is favored at low T and high P, while desorption occurs at higher T and lower P.

Typical practical conditions correspond to adsorption at 30 atm and 25 °C, and desorption at 3 atm and 100 °C [37]. The corresponding storage capacities are summarized in Table 2. While the $C_T$ obtained from DFT reaches 13.08, 10.82, 9.23, and 9.15 wt%, the effective reversible capacities ($C_E$) under practical conditions are reduced to 8.06, 4.26, 2.05, and 7.85 wt% for Li-, Na-, K-, and Ca-decorated N-GDY, respectively. Therefore, the Li- and Ca-decorated N-GDY maintain effective gravimetric capacities above the DOE target value, highlighting them as particularly promising candidates for practical $H_2$ storage applications.

Finally, Table 3 compares the performance of the present systems with previously reported hydrogen storage materials based on $N_T$, $E_{ads}$, and $C_T$. The results demonstrate that light-metal-decorated N-GDY monolayers exhibit competitive or superior performance compared with many other proposed 2D materials, confirming their potential as efficient hydrogen storage platforms.

**IV. Conclusion**

In this work, State-of-the-art computational methods based on DFT, AIMD, and thermodynamic analysis were utilized to investigate $H_2$ storage on light-metal-decorated N-GDY monolayers. Four light metals (Li, Na, K, and Ca) were examined as dopants to evaluate their stability and hydrogen adsorption capability. The calculations demonstrate that up to five metal atoms can be stably embedded within the pores of the N-GDY lattice, as confirmed by $E_{bind}$ exceeding the corresponding bulk $E_{coh}$ and by AIMD simulations at 400 K. Metal decoration induces metallization of the N-GDY monolayer and results in significant charge transfer from the metal atoms to the carbon–nitrogen framework, generating positively charged metal centers that effectively polarize $H_2$ molecules and enhance physisorption interactions. $H_2$ adsorption simulations show that each metal atom can adsorb approximately five $H_2$ molecules with average adsorption energies ranging from −0.15 to −0.22 eV per molecule, which falls within the desirable range for reversible hydrogen storage. The resulting intrinsic gravimetric storage capacities reach 13.08, 10.82, 9.23, and 9.15 wt% for Li-, Na-, K-, and Ca-decorated N-GDY, respectively, exceeding the DOE ultimate system-level target of 6.5 wt%. Thermodynamic analysis under practical adsorption and desorption conditions indicates that Li- and Ca-decorated N-GDY retain effective gravimetric capacities above the DOE benchmark. These findings highlight light-metal-functionalized N-GDY as a promising platform for high-capacity and reversible hydrogen storage.

**CRediT Authorship Contribution Statement**

**Wael Othman:** Investigation, Formal analysis, Methodology, Visualization, Validation, Project administration, Software, Writing - original draft preparation. **Ibrahim Al Ghoul:** Investigation, Formal analysis, Methodology, Validation, Writing- Reviewing and Editing. **Nacir Tit:** Conceptualization, Investigation, Formal analysis, Methodology, Resources, Writing- original draft preparation. **K-F. Aguey-Zinsou**: Funding acquisition, Resources, Supervision. **Tanveer Hussain:** Conceptualization, Funding acquisition, Resources, Supervision, Writing- Reviewing and Editing.

**Declaration of Competing Interest**

The authors declare that they have no known competing financial interests or personal relationships that could have appeared to influence the work reported in this paper.

**Acknowledgment**

This research was conducted using high-performance computing resources at Khalifa University, New York University Abu Dhabi, and United Arab Emirates University. The authors acknowledge funding from the ARC Training Centre for the Global Hydrogen Economy (GlobH2E, IC200100023). Computational resources were provided by the National Computational Infrastructure (NCI), an NCRIS-enabled capability supported by the Australian Government. Access to these resources was facilitated through the Sydney Informatics Hub HPC Allocation Scheme, supported by the Deputy Vice-Chancellor (Research) at the University of Sydney.

# Figures

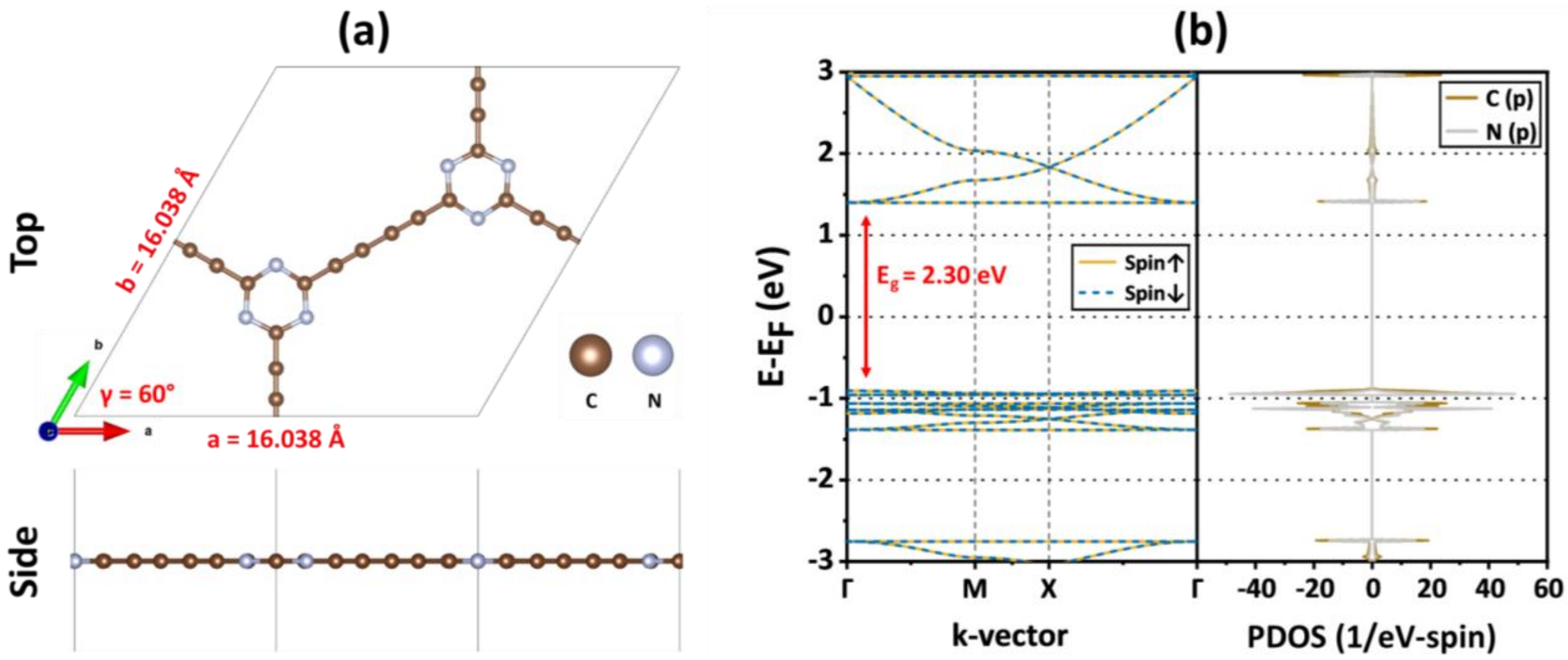

**Figure 1.** (a) Optimized atomic structures (top and side views) of pristine nitrogen–graphdiyne (N-GDY) monolayer. The primitive cell contains 18 carbon and 6 nitrogen atoms arranged in a hexagonal structure with lattice parameters a = b = 16.038 Å and c = 20 Å, α = β = 90°, and γ = 60°. (b) Spin-polarized band structure and projected density of states (PDOS) of pristine N-GDY, showing a direct band gap of 2.30 eV with paramagnetic characteristics. High symmetry k-points are M = 0.2262 $Å^{-1}$, X = 0.3568 $Å^{-1}$, and Γ = 0.6180 $Å^{-1}$. The Fermi level (horizontal dashed line) is shifted to zero.

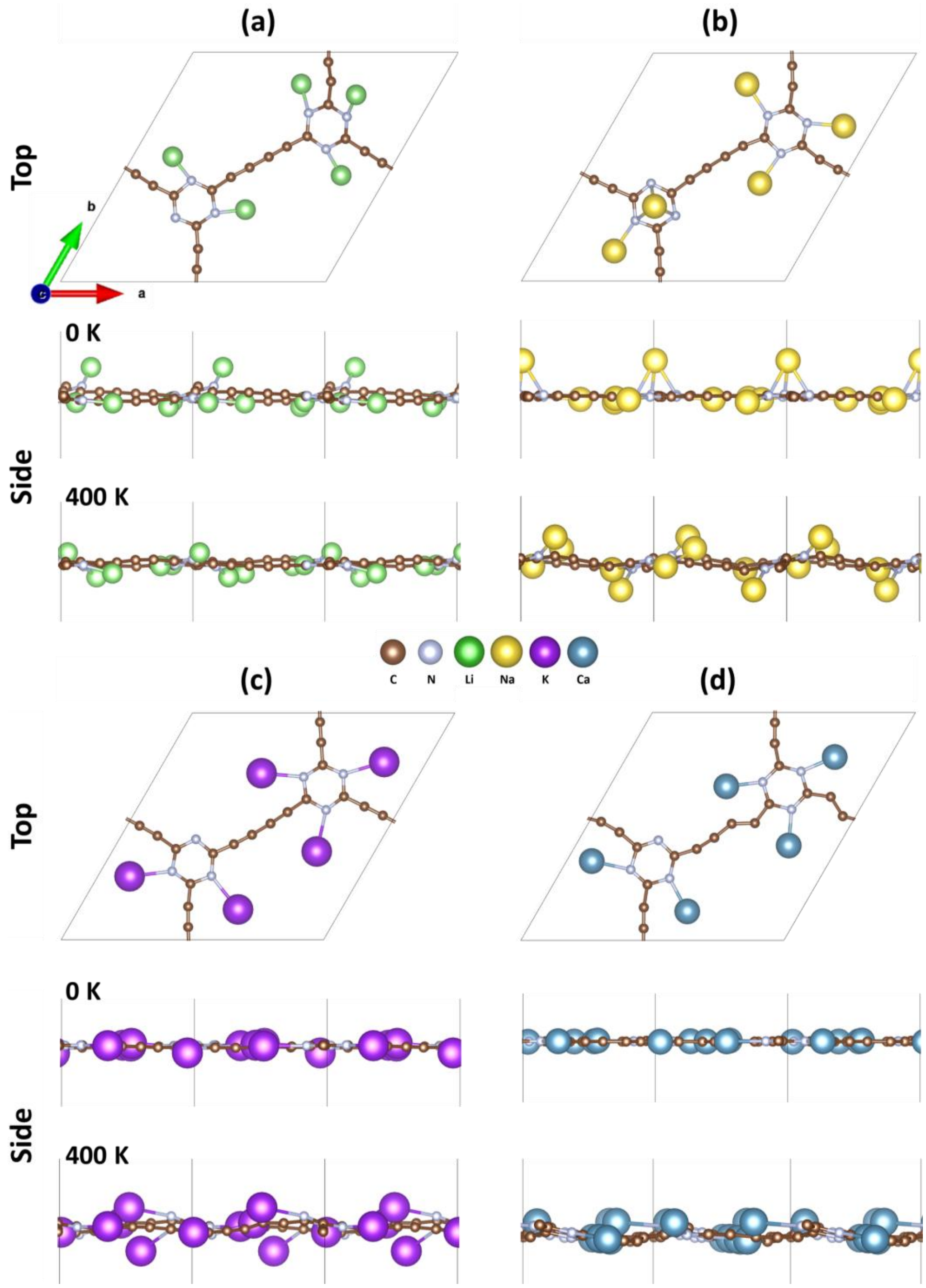

**Figure 2.** Optimized atomic structures (top and side views) of (a) 5Li@N-GDY, (b) 5Na@N-GDY, (c) 5K@N-GDY, and (d) 5Ca@N-GDY. The light-metal dopants act as active sites for $H_2$ adsorption. AIMD simulations at 400 K confirm the thermodynamic stability of the light-metal-functionalized N-GDY monolayers.

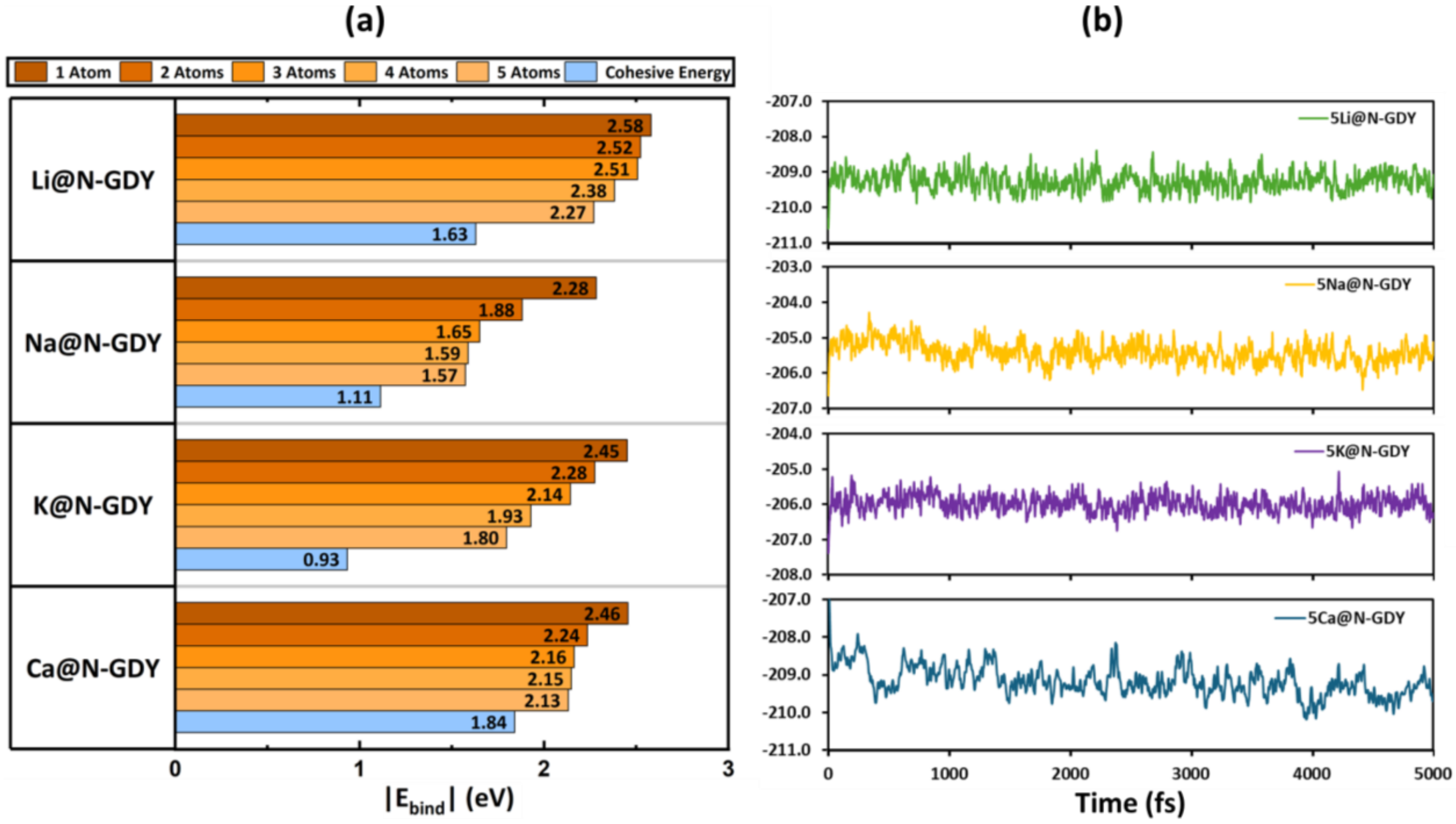

**Figure 3.** (a) Average binding energy ($|E_{bind}|$) per light-metal dopant on N-GDY (up to 5 atoms per primitive cell), compared to the corresponding cohesive energy of each bulk metal. (b) AIMD simulation results at 400 K for light-metal (Li, Na, K, and Ca) functionalized N-GDY, indicating robust thermodynamic stability.

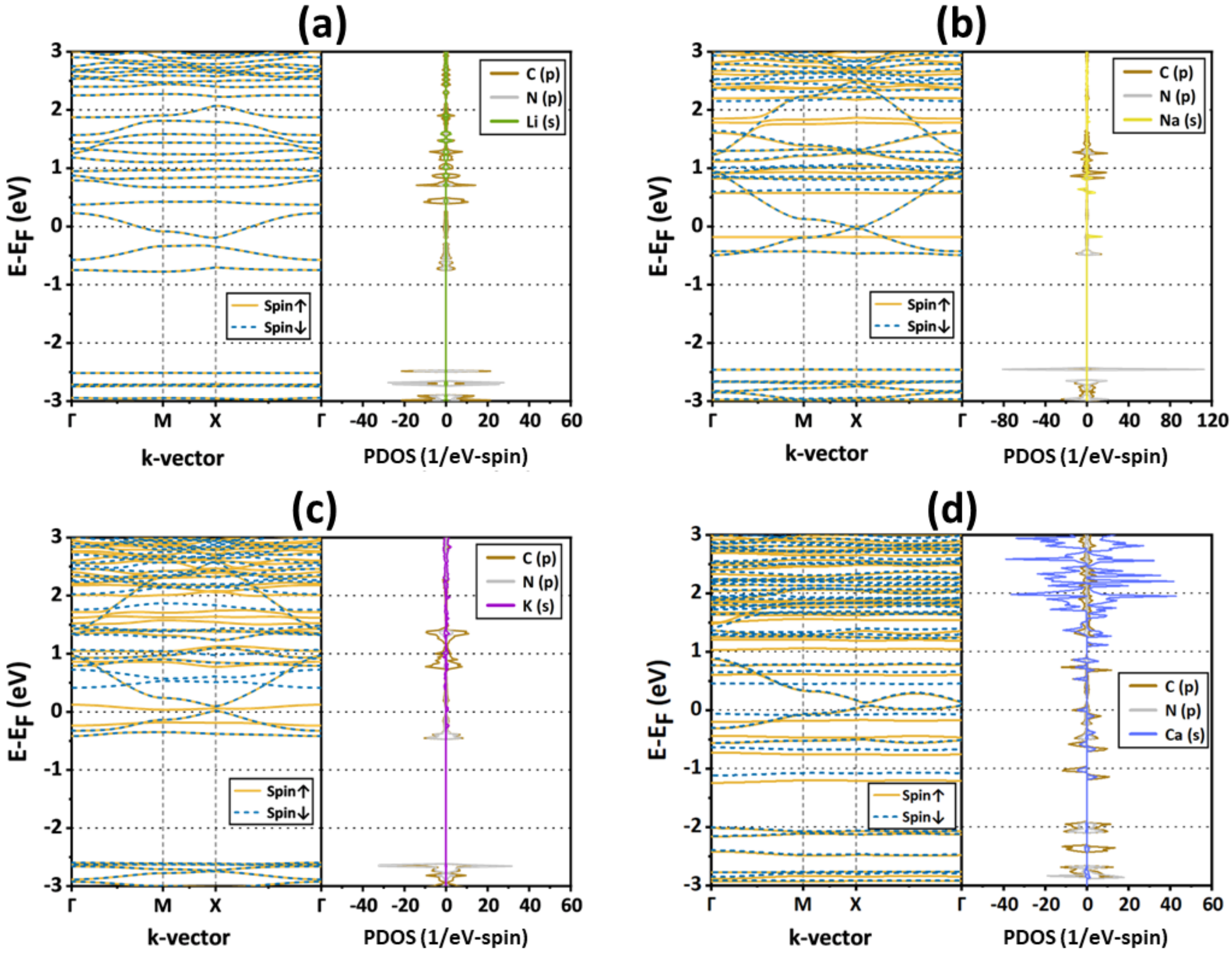

**Figure 4.** Spin-polarized band structure and projected density of states (PDOS) of (a) 5Li@N-GDY, (b) 5Na@N-GDY, (c) 5K@N-GDY, and (d) 5Ca@N-GDY. The k-points M, X, and Γ correspond to 0.2262, 0.3568, and 0.6180 $Å^{-1}$, respectively. The Fermi level (horizontal dashed line) is shifted to zero.

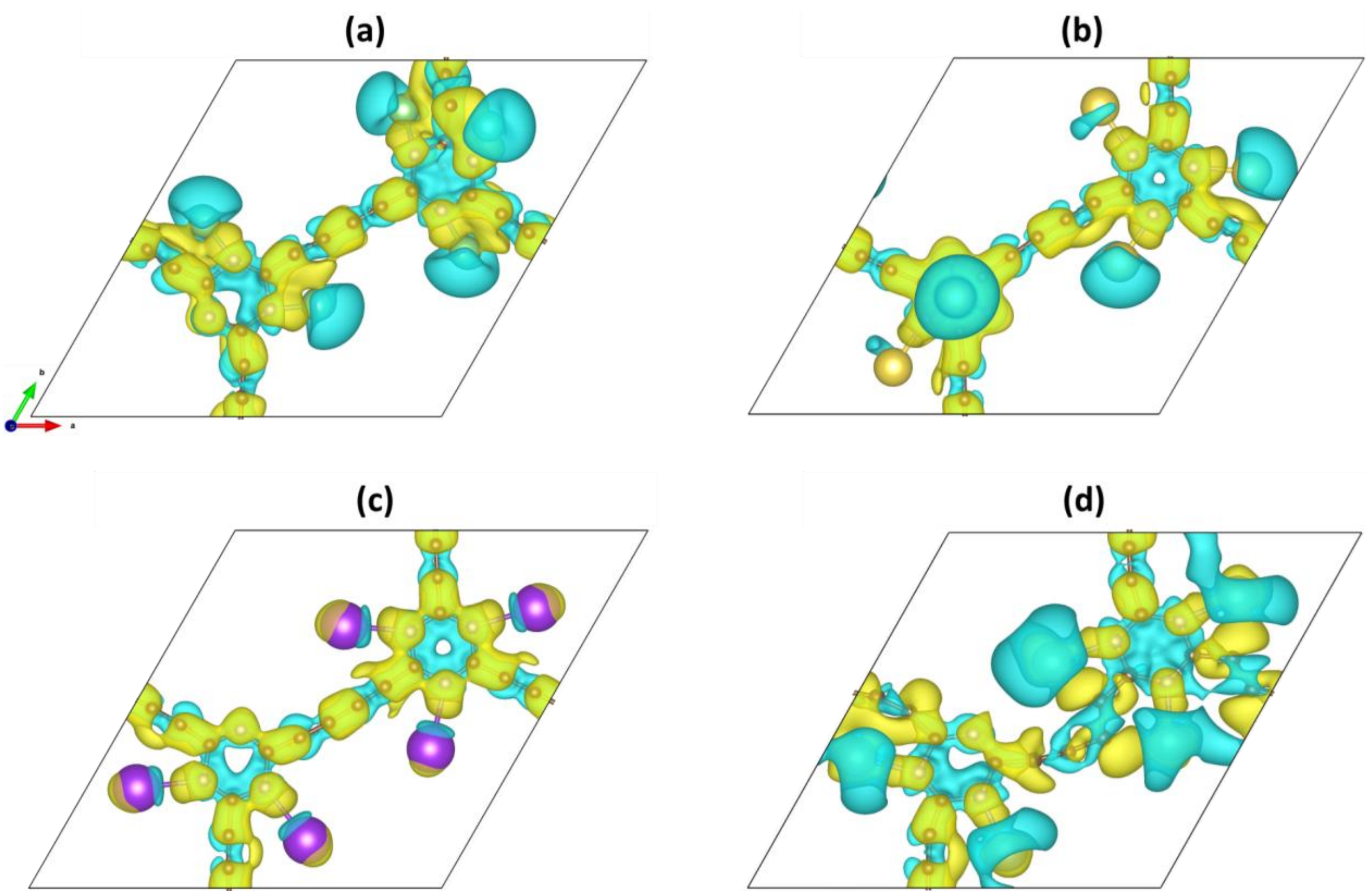

**Figure 5.** Charge density difference (CDD) plots showing the bond characteristics based on maximally localized Wannier functions (MLWFs) for (a) 5Li@N-GDY, (b) 5Na@N-GDY, (c) 5K@N-GDY, and (d) 5Ca@N-GDY. Yellow iso-surfaces represent regions of charge accumulation, while cyan iso-surfaces indicate charge depletion. The iso-surface value is set to 0.0018 e/Bohr$^3$.

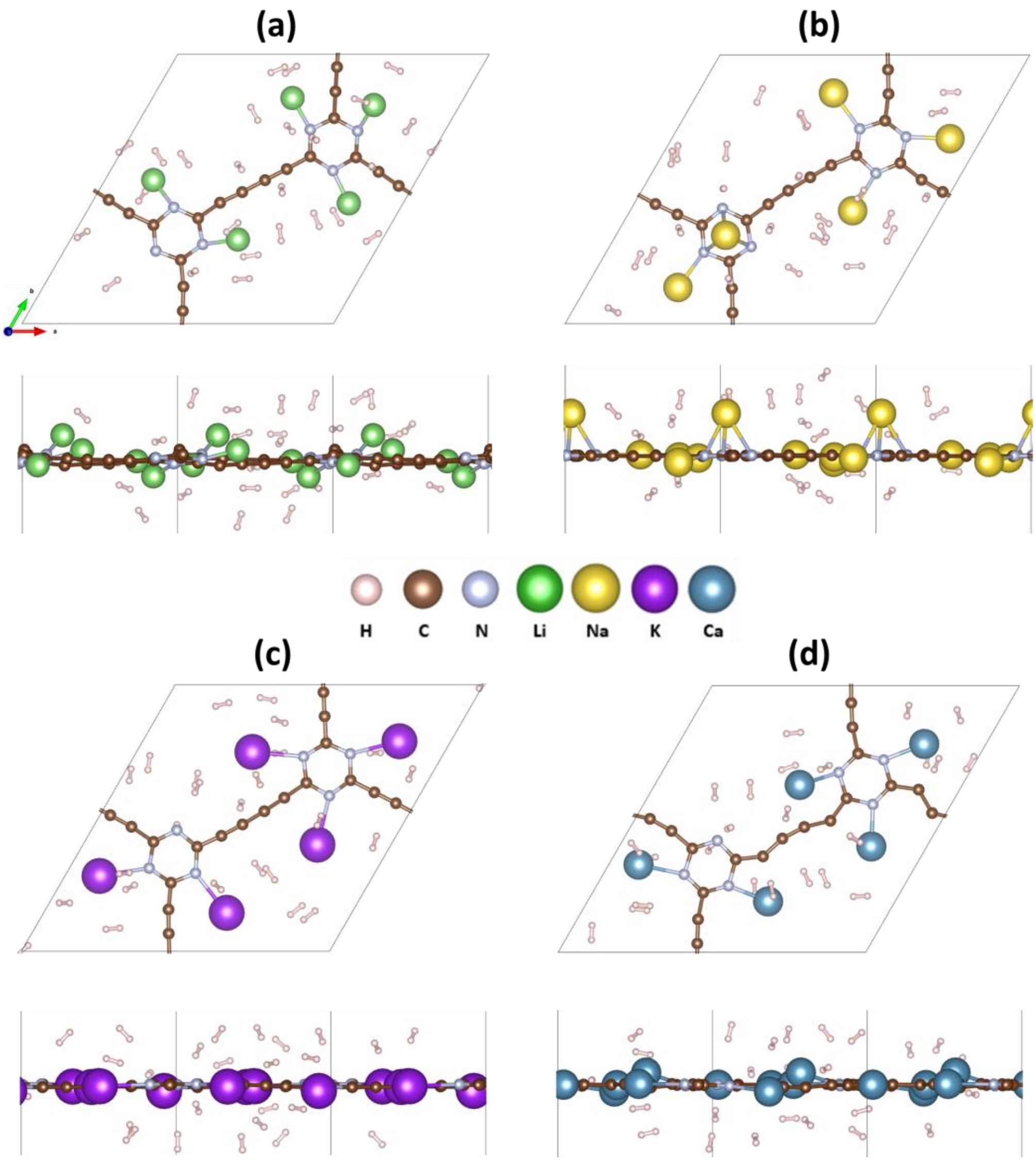

**Figure 6.** Optimized atomic structures (top and side views) of fully hydrogenated light-metal-functionalized N-GDY monolayers. (a) 5Li@N-GDY with 25 $H_2$ molecules, (b) 5Na@N-GDY with 25 $H_2$ molecules, (c) 5K@N-GDY with 25 $H_2$ molecules, and (d) 5Ca@N-GDY with 25 $H_2$ molecules.

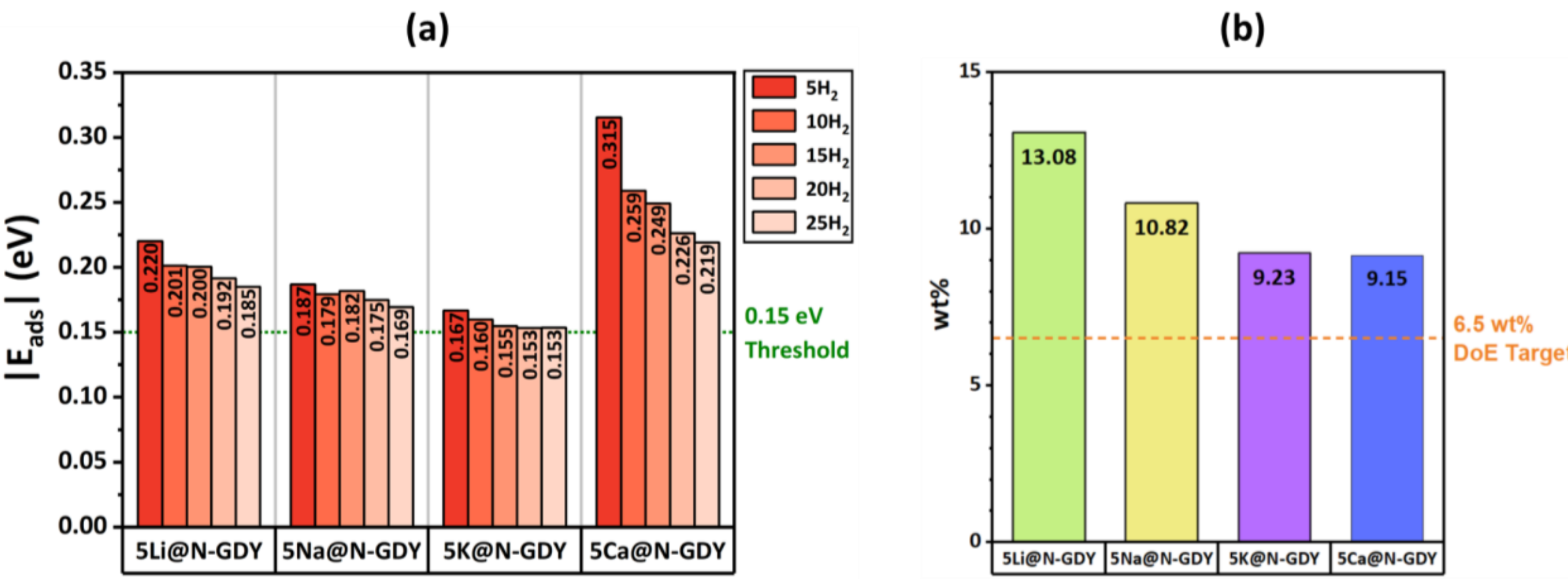

**Figure 7.** Evaluation of fully hydrogenated light-metal-doped N-GDY monolayers. (a) Average adsorption energy per $H_2$ molecule for each system, all exceeding the optimal lower threshold of 0.15 eV. (b) Corresponding theoretical gravimetric hydrogen storage capacities (wt%) under maximum hydrogenation, all surpassing the DOS's ultimate target of 6.5 wt%.

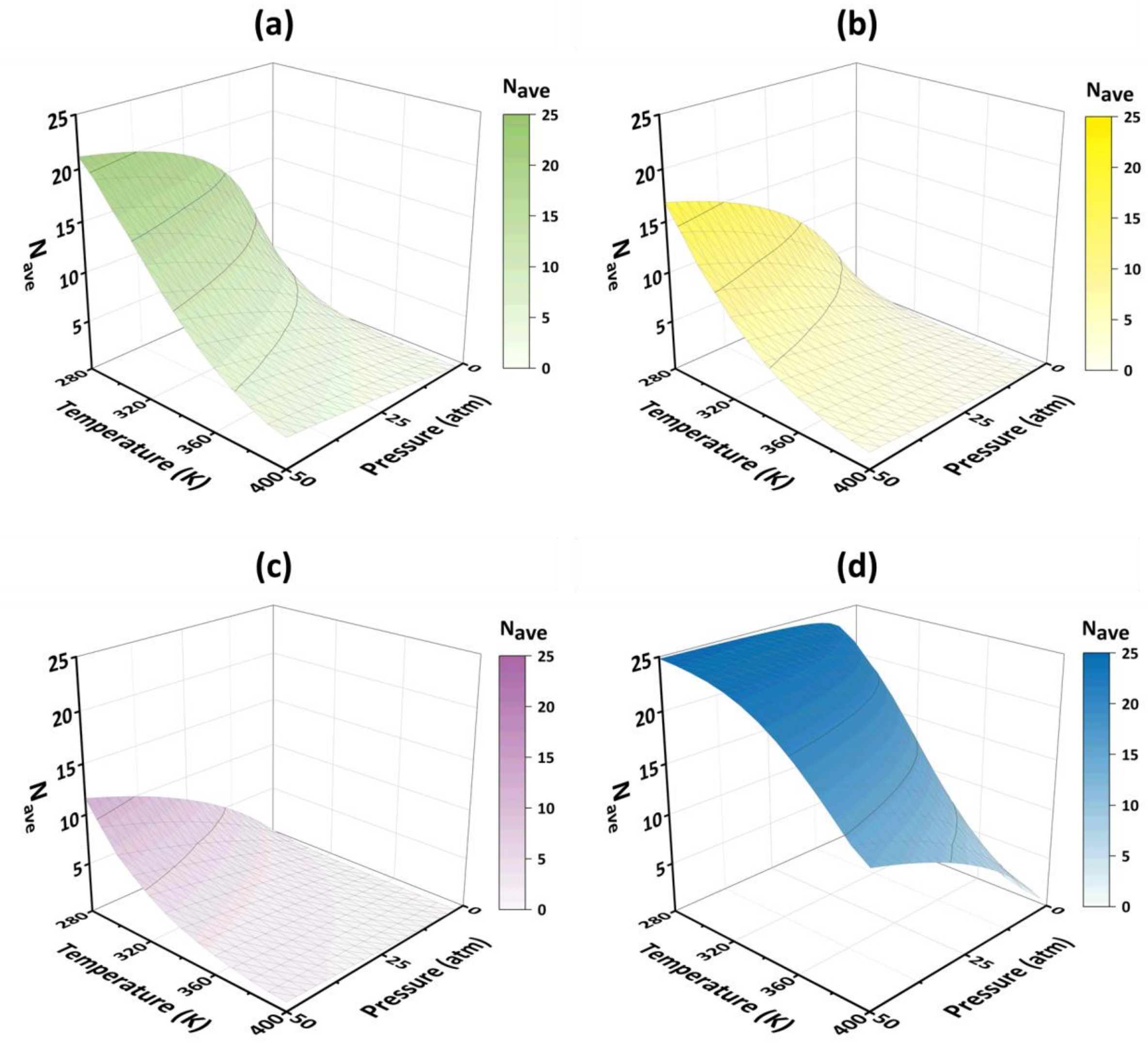

**Figure 8.** Thermodynamic analysis showing the average number of adsorbed $H_2$ molecules ($N_{ave}$) as a function of temperature and pressure for (a) 5Li@N-GDY, (b) 5Na@N-GDY, (c) 5K@N-GDY, and (d) 5Ca@N-GDY.

**Table 1.** Electronic properties of pristine and light-metal-functionalized N-GDY monolayers. Δq, $E_F$, $E_g$, and M represent the Bader charge transfer, Fermi level, band gap, and total magnetization, respectively.

| System | Δq (e/MI) | $E_F$ (eV) | $E_g$ (eV) | M ($\mu_B$) |
|---|---|---|---|---|
| Pristine N-GDY | NA | -5.71 | 2.30 (↑)(↓) | 0 |
| 5Li@N-GDY | 4.386 | -2.02 | 0.02 (↑)(↓) | 0 |
| 5Na@N-GDY | 3.520 | -2.00 | 0.0050 (↑)<br>0.0057 (↓) | 0.93 |
| 5K@N-GDY | 3.357 | -1.84 | 0.0033 (↑)<br>0.0120 (↓) | 1.22 |
| 5Ca@N-GDY | 5.331 | -2.33 | 0.0058 (↑)(↓) | 1.17 |

**Table 2.** Theoretical and practical $H_2$ storage capacities of light-metal-functionalized N-GDY monolayers. The theoretical number of adsorbed $H_2$ molecules ($N_T$) is obtained from DFT simulations and used to calculate the theoretical gravimetric storage capacities ($C_T$) using equation (3). Classical thermodynamic analysis is employed to estimate the number of adsorbed $H_2$ molecules under adsorption ($N_a$: P = 30 atm, T = 25°C) and desorption ($N_d$: P = 3 atm, T = 100°C) conditions. The practical number of reversibly adsorbed/released $H_2$ molecules ($N_p = N_a - N_d$) is used to calculate the effective storage capacity ($C_E$).

| Storage material | $C_T$ (wt%) | $N_T$ (molecule) | $N_a$ (molecule) | $N_d$ (molecule) | $N_p$ (molecule) | $C_E$ (wt%) |
|---|---|---|---|---|---|---|
| 5Li@N-GDY | 13.08 | 25 | 14.93 | 0.36 | 14.57 | 8.06 |
| 5Na@N-GDY | 10.82 | 25 | 9.35 | 0.18 | 9.17 | 4.26 |
| 5K@N-GDY | 9.23 | 25 | 5.23 | 0.09 | 5.14 | 2.05 |
| 5Ca@N-GDY | 9.15 | 25 | 24.18 | 3.03 | 21.15 | 7.85 |

**Table 3.** Comparison of various hydrogen storage systems based on theoretical number of adsorbed $H_2$ molecules per simulation cell ($N_T$), average adsorption energy per $H_2$ molecule ($E_{ads}$), and theoretical $H_2$ gravimetric capacity ($C_T$). The computational methods employed in each study are indicated.

| **Hydrogen Storage System** | **$N_T$** | **$E_{ads}$ (eV)** | **$C_T$ (wt%)** |
|---|---|---|---|
| Li-decorated N-GDY* (this work) | 25 | -0.19 | 13.08 |
| Na-decorated N-GDY* (this work) | 25 | -0.17 | 10.82 |
| K-decorated N-GDY* (this work) | 25 | -0.15 | 9.23 |
| Ca-decorated N-GDY* (this work) | 25 | -0.22 | 9.15 |
| Li-decorated $C_9N_4$* [38] | 6 | -0.20 | 11.90 |
| V-decorated 2DPA-I** [39] | 7 | -0.44 | 7.29 |
| V-decorated biphenylene* [40] | 7 | -0.47 | 10.30 |
| Ti-decorated graphene** [41] | 8 | -0.42 | 7.80 |
| Ti-doped ψ-graphene* [42] | 9 | -0.30 | 13.41 |
| Li-decorated B@$r_{57}$haeckelite* [43] | 12 | -0.16 | 10.00 |
| Sc-doped Holey graphyne*** [44] | 5 | -0.36 | 9.80 |
| Ca-decorated B-doped siligene** [45] | 7 | -0.20 | 13.79 |
| Li-decorated B-doped siligene** [46] | 4 | -0.17 | 12.71 |
| Li-decorated $C_2N$** [35] | 20 | -0.14 | 13.00 |

| Li-decorated borophene** [47] | 14 | -0.20 | 13.96 |
|---|---|---|---|
| Li-decorated graphyne** [48] | 12 | -0.26 | 18.60 |
| Li-decorated Holey graphyne** [49] | 24 | -0.22 | 12.80 |
| Ti-decorated $C_2N$* [50] | 10 | -0.28 | 6.80 |
| Zr-decorated biphenylene* [51] | 9 | -0.40 | 9.95 |
| Li-decorated biphenylene***** [52] | 12 | -0.20 | 7.40 |
| K-decorated biphenylene** [53] | 5 | -0.24 | 11.90 |
| Li-decorated $B_2S$ honeycomb* [54] | 12 | -0.14 | 9.10 |
| Y-decorated $C_{48}B_{12}$** [55] | 72 | -0.46 | 7.51 |
| Li-decorated MOF-5** [56] | 18 | _ | 4.30 |
| Y-decorated covalent triazine frameworks* [57] | 7 | -0.33 | 7.30 |
| Sc-decorated g-$C_3N_4$** [58] | 7 | -0.39 | 8.55 |
| Ti-decorated graphene**** [59] | 8 | -0.46 | 6.11 |
| Y-decorated g-$C_3N_4$** [60] | 9 | -0.33 | 8.55 |
| Ti-decorated B-doped twin-graphene* [61] | 8 | -0.20 | 4.95 |
| Co-decorated N-doped graphene** [62] | 28 | -0.19 | 11.36 |
| V-decorated porous graphene**** [63] | 6 | -0.56 | 4.58 |
| Ti-decorated graphene** [64] | 8 | -0.21 | 6.30 |
| Na-decorated siligene** [65] | 32 | -0.16 | 14.16 |
| K-doped $PC_{71}BM$** [66] | 45 | -0.14 | 6.22 |
| Ca-decorated DCV graphene**** [67] | 14 | -0.10 | 5.80 |
| Li-decorated graphene nanoribbons** [68] | 8 | -0.24 | 3.80 |
| Li-decorated $T_{4,4,4}$-graphyne* [69] | 16 | -0.20 | 10.46 |
| K-decorated Ga-doped germanene** [70] | 36 | -0.19 | 8.19 |
| Li-decorated P-$BN_2$** [71] | 28 | -0.16 | 13.27 |
| Li-decorated N-doped penta-graphene** [72] | 12 | -0.24 | 7.88 |
| Li-decorated defective penta-$BN_2$** [73] | 16 | -0.14 | 9.17 |
| Li-decorated $B_3S$* [74] | 12 | -0.17 | 7.70 |
| Li-decorated penta-silicene**** [75] | 12 | -0.22 | 6.42 |
| Li-decorated penta-octa-graphene* [76] | 3 | -0.22 | 9.90 |
| γ–graphyne**** [77] | 69 | -0.20 | 13.90 |
| K-decorated SnC** [78] | 6 | -0.20 | 5.50 |
| Li-decorated borophene**** [79] | 10 | -0.36 | 9.00 |
| Li-decorated boron phosphide** [80] | 16 | -0.19 | 7.40 |
| Li-decorated defected biphenylene* [81] | 14 | -0.20 | 8.75 |
| Li-decorated honeycomb borophene oxide** [82] | 32 | -0.24 | 8.3 |
| Li-decorated honeycomb borophene oxide** [83] | 8 | -0.22 | 9.84 |
| Li-decorated $B_4N$** [84] | 16 | -0.16 | 6.23 |

*GGA-PBE/DFT-D3

**GGA-PBE/DFT-D2

***GGA-PBE/DFT-D2 and HSE06

****GGA-PBE/DFT-D

*****VDW-DF/DZP